\documentclass{cimento1}
\usepackage{graphicx}
\usepackage{amssymb}
\usepackage{epstopdf}

\newcommand{\qtop}[0]{\ensuremath{t\bar t}~ }
\newcommand{\met}[0]{\ensuremath{E_{T}^{\mathrm{miss}}}~ }
\newcommand{\Zee}[0]{\ensuremath{Z\rightarrow ee }~}
\newcommand{\Rlj}[0]{\ensuremath{R_{l-\mathrm{jets}}}}
\title{Commissioning ATLAS and CMS with top quarks}
\author{
	B. S. ~Acharya\from{i1}
	F. ~Cavallari\from{i2},
	G. ~Corcella\from{i5}\from{i51},
	R. ~Di Sipio\from{i4} \atque
	G. ~Petrucciani\from{i51}}

\instlist{ 
	\inst{i1} Abdus Salam International Center for Theoretical Physics and INFN, Sezione di Trieste, Italy
	\inst{i2} INFN, Sezione di Roma, Italy
	\inst{i5} Museo Storico della Fisica e Centro Studi e Ricerche E.~Fermi, Roma, Italy
	\inst{i51} Scuola Normale Superiore and INFN, Sezione di Pisa, Italy
 	\inst{i4} Universit\`a di Bologna and INFN, Sezione di Bologna, Italy
}

\PACSes{
\PACSit{14.65.Ha }{Top quarks} 
\PACSit{13.38.Be }{Decays of $W$ bosons} 
\PACSit{13.85.Hd}{Inelastic scattering: many-particle final states} 
}

\begin{document}
\maketitle
\begin{abstract}
The large \qtop production cross-section at the LHC suggests the use of top quark decays to calibrate several critical parts of the detectors, such as the trigger system, the jet energy scale and $b$-tagging. 
\end{abstract}


\section{Introduction}

Events in which top-quark pairs are produced will be extremely important at the LHC,
as they will provide a unique environment to study physics
within the Standard Model and beyond \cite{beneke}.
Final states in $t\bar t$ events are classified in three categories, according to the
$W$-decay mode in top decay $t\to bW$:
fully hadronic, semileptonic or fully leptonic.
Semileptonic \qtop decays produce complex 
signatures within the detector, involving missing transverse energy, charged leptons,
light-particle jets 
and $b$-jets. 
Therefore, in order to study these events accurately at the LHC, 
the understanding of all the parts of the detectors is mandatory. 
In particular,
the following  should be mastered:
\begin{itemize}
\item Trigger system;
\item Lepton and jet reconstruction;
\item Calculation of missing transverse energy;
\item $b$-tagging. 
\end{itemize}
Conversely, the top quark is an excellent instrument, 
thanks to 
the large $t\bar t$ 
cross-section at the LHC, $\sigma(t\bar t) \sim 830$~pb, 
more than 100 times larger than at the Tevatron accelerator
\cite{bonciani}. Semileptonic \qtop events
link all these items together and can therefore be used to make
what is commonly referred to as an \emph{in-situ} calibration.

\section{Triggers}
At the LHC collisions will happen with a frequency of up to 40 MHz, and this number has to be compared with the capabilities of the ATLAS and CMS mass storage 
systems of about 200 Hz. So, the trigger system of ATLAS and CMS has been designed to select one event out of 10 millions, when running at the highest design luminosity.


The ATLAS and CMS trigger systems both have a hardware-based level 1 and a software-based high-level trigger\cite{ref:cms:redisc,ref:atlas:trig}. 
Level 1 makes use of the muon detectors and calorimeters in order to identify particles, while higher levels perform a more refined reconstruction. 
In order to choose interesting events and reduce the output rate, the
two experiments designed their trigger systems in two different ways.

ATLAS makes use of \emph{Regions of Interest} (RoI), a technique that gives access to high-granularity information only for the regions flagged as interesting by the Level 1 Trigger. The CMS High Level Trigger can access the full detector readout, but it performs only the minimal amount of reconstruction needed to determine if an event has to be accepted or dropped. At the end of the process, both ATLAS and CMS trigger systems will write data with a frequency of about 100 Hz and a latency of few $\mu s$.

\begin{figure}[hp]
\begin{center}
    \includegraphics[width=0.4\textwidth]{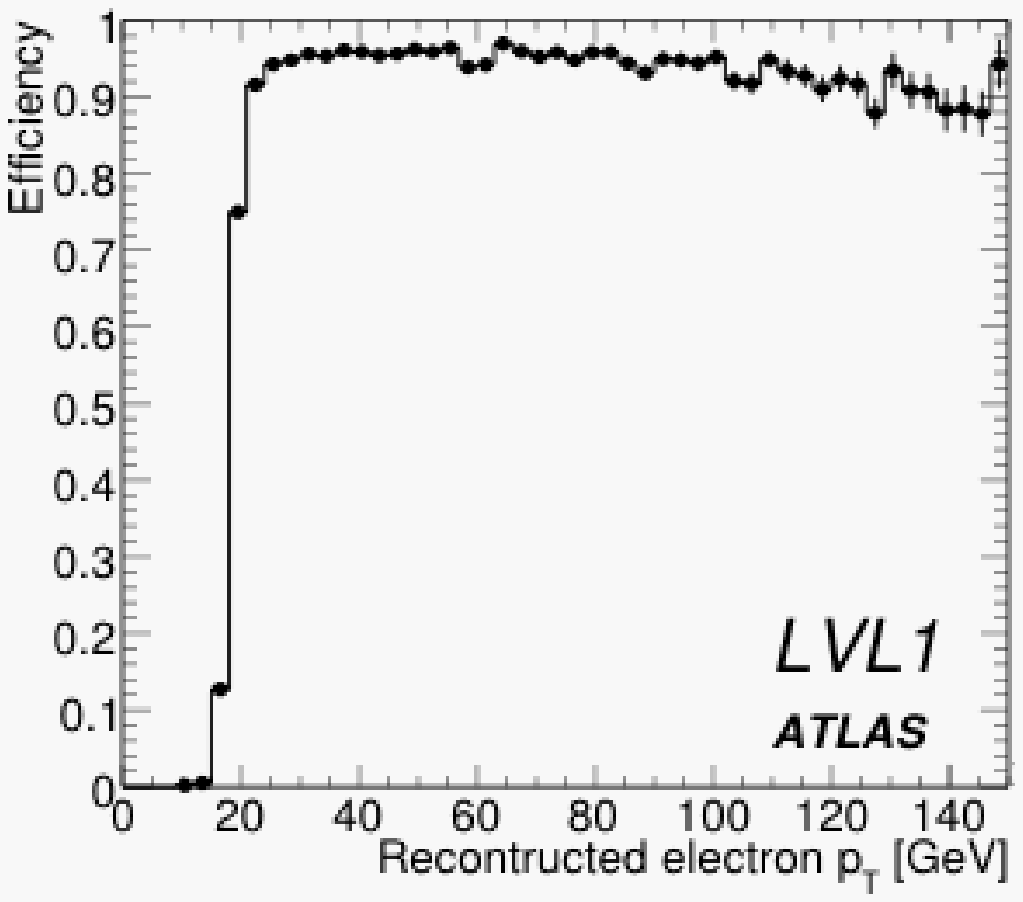}
    \includegraphics[width=0.4\textwidth]{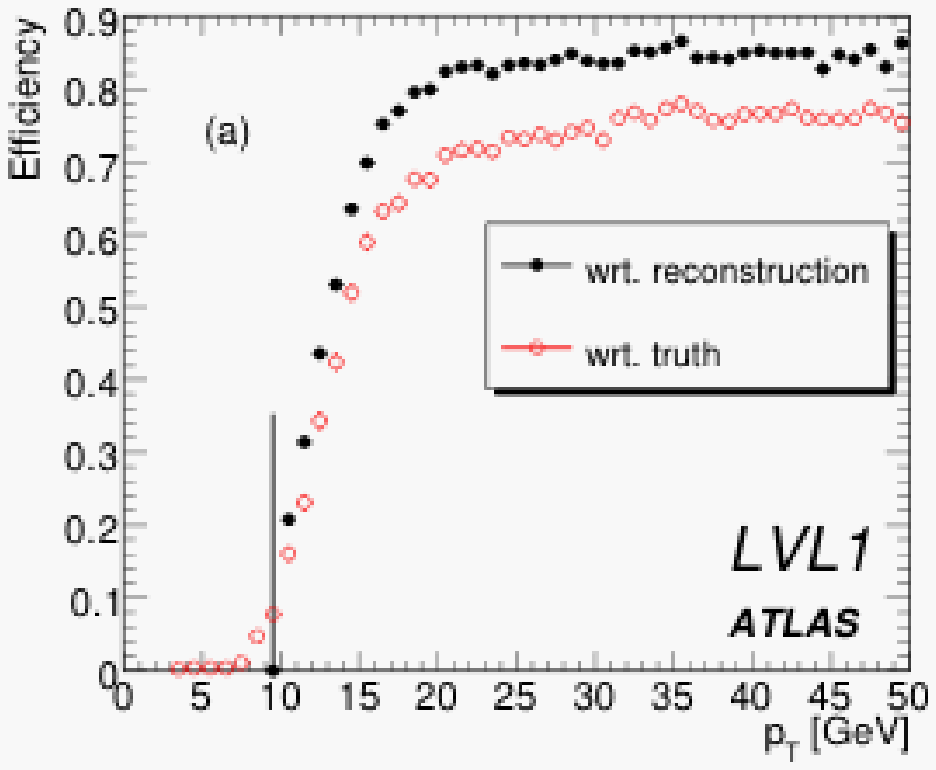}
  \caption[ATLAS L1 1-lept trigger]{\small{ATLAS 
Level-1 efficiencies for one-lepton trigger menu e251 (left) and mu20i (right). Efficiencies are calculated with respect to the offline reconstruction. }}
  \label{fig:trig_atlas_lvl1emu}
\end{center}
\end{figure}

In the first days of data-taking, close attention will be paid to the study of the single-lepton triggers. 
In fact, a large number of important processes involve the production of at least one isolated 
charged lepton, and leptonic decays of top quarks are amongst these. 
Fig.~\ref{fig:trig_atlas_lvl1emu} shows the efficiency of the level-1 single-lepton triggers in \qtop events, 
calculated with respect to the offline reconstruction.   

Moreover, semileptonic events are triggered
from jet triggers, too, giving the possibility to measure directly the efficiency of the leptonic triggers.
Thus, the very large cross section of \qtop events can be successfully exploited to calibrate the triggers. 

For example, 
a sample of events can be collected according to the offline selection
defined in Tab. \ref{tab:tt_offline_sel}.
Then, a sub-sample is extracted, containing only events that fired the single-lepton trigger. From this sub-sample, 
one can easily calculate the fraction of \qtop events that fired the jet trigger. 
This technique can be subsequently applied, 
e.g., to several jet triggers for each lepton trigger, 
leading to a very good determination of combined trigger efficiencies \cite{ref:atlas:trig}.

Top-quark production is also suitable to study other triggers, such as double-lepton (for full-leptonic decays), 
jet and missing-$E_{T}$ triggers. In fact, 
two leptons give a very clean signature 
for triggering, albeit limited in statistics at the very beginning. 

With early data, the fully hadronic channel is extremely challenging triggerwise, due to the large QCD background. 
Reasonably, this channel will be studied accurately in a subsequent phase of the 
experiment. 

ATLAS and CMS will also estimate the single-lepton 
trigger efficiency as a function of its momentum from processes which do not involve top quarks, such as
\Zee / $\mu\mu$. 
However, since the
jet energy scale and underlying event might be different between \Zee and \qtop 
processes, it is clearly preferable 
to calculate efficiencies for \qtop events by using \qtop events themselves.


\section{Jet Energy Scale}
The cone algorithm for jet reconstruction, with a cone radius $R=0.4$ or 0.5, 
is commonly used 
both in ATLAS and CMS, since it provides a good compromise 
between energy reconstruction and angular resolution \cite{ref:atlas:ljets}.
 
Due to a HCAL  resolution lower than ATLAS, CMS found better results using the
\emph{particle flow}, a useful technique when dealing with low-granularity calorimeters. Preliminary studies show an overall efficiency 
similar to that of ATLAS.

The (mis)calibration of the Jet Energy Scale (JES) appears as an important source of systematic uncertainty 
on $M_{W}$ and $M_{t}$. The \emph{a priori} knowledge of jet-energy calibration is about $10\%$. 
The goal of 1~GeV error on $M_t$  requires understanding the JES to
1\%. The Jet Energy Scale can be evaluated using  the method of $p_T$-balance applied to 
$Z/\gamma$+jets events.
Here, the well reconstructed Z/$\gamma$ transverse momentum
can be balanced against the jets in the events, allowing a $p_T$-dependent jet calibration.
These processes are also useful for the $b$-jet energy scale, when jets are tagged as $b$-jets.

However, as stated before, it would be better to measure the JES for \qtop events by means of \qtop events 
themselves, at least for two main reasons: 
\begin{itemize}
\item \qtop selection cuts can lead to JES different from that of $Z/\gamma$+jets; 
\item the underlying event (UE) may be different for the two processes. 
\end{itemize}

To this end one could exploit the $t\to Wb\to jjb$ decay chain, since it gives an identifiable $W\to jj$ sample (\emph{in-situ} 
calibration). The $jj$ invariant mass should of course yield the well-known $W$-boson mass.

ATLAS will determine the JES by studying the reconstructed $M_{W}$ 
after the offline selection (as defined in Tab \ref{tab:jes_offline_sel}). 
Its impact, after varying the reconstructed jet energies by $\pm 1\%$, 
has been evaluated.
Unfortunately, offline selection introduces a bias caused by the $p_{T}$-cut which is
important near threshold \cite{ref:atlas:ljets}. 
To handle this problem, fitting techniques are applied to determine
the jet-energy-scale factors as a function of the jet energy and pseudorapidity $\eta$.

\begin{figure}[hp]
\begin{center}
    \includegraphics[width=0.3\textwidth]{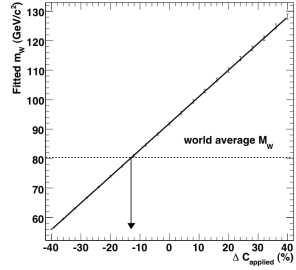}
  \caption[CMS JES M_{W} fit]
{\small{CMS will calculate JES by fitting the reconstructed $M_{W}$ to the world average. 
Corrections are applied by multiplying $M_{W}^{\mathrm{reco}}$ by a constant term $\Delta C$.}}
  \label{fig:jes_cms}
\end{center}
\end{figure}

Starting from the same principles, CMS will calculate $M_{W}$ by combining the two jets.
The light-quark jet energy is scaled by a global correction factor $\Delta C$, chosen to fit the reconstructed
$W$ mass within the  world average, 
as shown in Fig.~\ref{fig:jes_cms}. Studies show that the main sources 
of systematic uncertainty on $\Delta C$
are the pile-up 
and $b$-tag efficiency \cite{ref:cms:jes}. More refined techniques are under study, based on a kinematic fit of $M_{bjj}$ in $t\to Wb\to jjb$ 
decays, so that one can also measure the $b$-JES.

\section{$b$-tagging}
Identification of $b$-jets is crucial in many analyses at the LHC, 
such as the searches for the Higgs boson, 
supersymmetry and other New Physics scenarios. Thus, to calibrate $b$-tagging algorithms, one would like to isolate  
a sample of $b$-jets as pure as possible.

Again, the large \qtop cross section offers the 
possibility of an \emph{in-situ} calibration with several 
advantages, since almost every \qtop event contains two $b$-quarks.
In fact, semileptonic \qtop events are identifiable {\it  without $b$-tagging} 
and hence give a handle on $b$-tagging mechanisms \cite{ref:atlas:btag}. With an integrated luminosity of 100/pb, 
several hundred events are expected. 
To gain more statistics, di-jet events could be used
but for b-tagging calibration. However the b-tagging efficiency, - like the JES - is sample and analysis dependent. For this reason a measurement of the efficiency from top events themselves is preferable.

\begin{figure}[hp]
\begin{center}
    \includegraphics[width=0.4\textwidth]{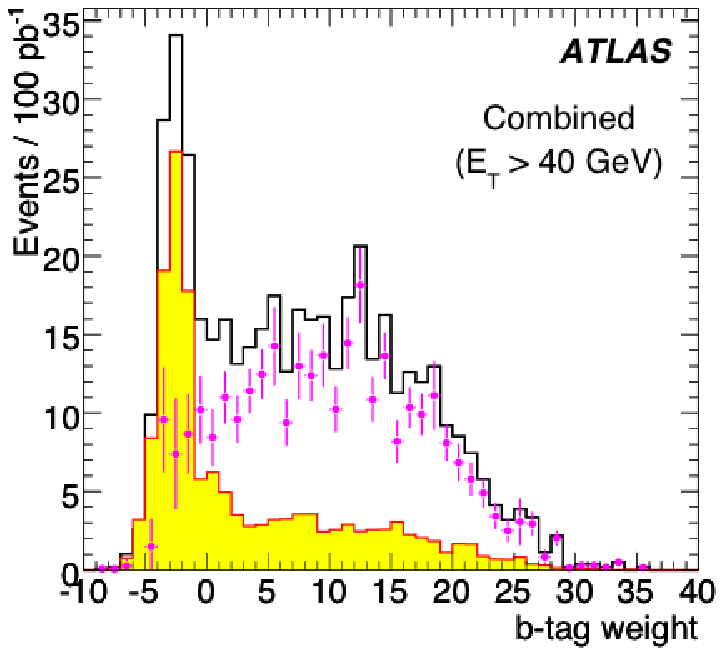}
    \includegraphics[width=0.4\textwidth]{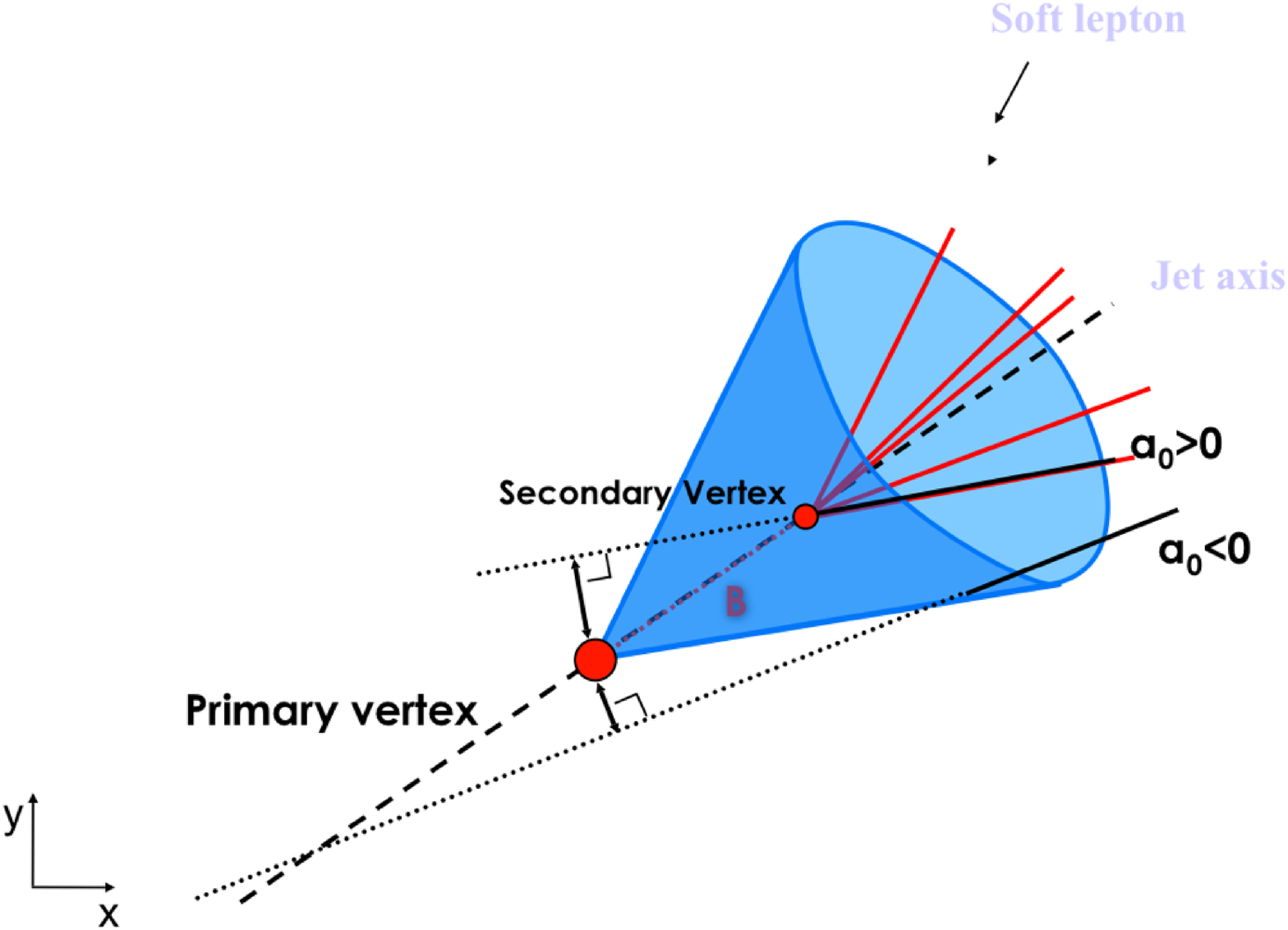}
  \caption[The ATLAS 
$b$-tag weight and CMS $b$-tag]
{\small{The ATLAS 
$b$-tagger 
weight $w$ (left) is low for light jets and high for $b$-jets. 
A generic selection is made applying a cut $w > 6$. CMS will make use of track counting and track probability taggers (right).}}
  \label{fig:btag_atlas_w}
\end{center}
\end{figure}

The default ATLAS $b$-tagger uses a likelihood algorithm weight $w$, 
constructed from the impact parameter
and the secondary-vertex taggers. Choosing a threshold on the weight translates into an efficiency to recognize correctly the $b$-jets 
($\epsilon_{b}$) and to reject the jets originated from the lighter quarks (\Rlj = $1/\epsilon_{l-jets}$).  
As shown in Fig. \ref{fig:btag_atlas_w}, $w$ is large for $b$-jets and low
for light-jets, proving itself as a good quantity to distinguish $b$-jets from light-jets. 
For example, setting the cut $w>6$, an overall efficiency $\epsilon_{b}=63\%$ and light-jet rejection \Rlj = 250
can be achieved \cite{ref:atlas:btag}.

In a recent CMS study an attempt was made to evaluate the
$b$-tagging efficiency directly from data.
In this study, a simple algorithm was used, 
mainly based on \emph{track counting} and \emph{track probability}. One can associate to each track inside the 
jet an impact parameter and a secondary
vertex location. If these values are greater than their thresholds, the track is `counted'.
Then, the jet is $b$-tagged if it contains more `counted' tracks than a minimum.  
Tagging one of the $b$-jets hardly allows to have a
rather pure $b$-jets sample from the other top
and evaluate the performance of the $b$-tagging.
However, this study shows that with 1fb$^{-1}$ one can reach an uncertainty on the $b$-tagging efficiency of $\sim$5\%.

Other strategies, such as the \emph{soft lepton} method, are also under study.

 

Assuming that each selected event actually contains two $b$-jets, $\epsilon_{b}$ can be measured from the data 
themselves, counting 
the number of tagged 
jets  as $b$-jets. To take into account mistagged events and backgrounds, a more refined 
likelihood function can be written, with $\epsilon_{b}$ and \Rlj~ as parameters.

Overall, the main sources of systematic uncertainty are light-jet rejection, JES, 
$W+$jets background contamination, and the uncertainty on 
the measurement of the top mass.

\section{Conclusions}
Top-quark events and, in particular, 
those with semileptonic decays, will be a poweful 
source of data for the measurement of trigger efficiencies, 
jet energy scale and $b$-tagging performance. 
Both ATLAS and CMS are developing methods and 
algorithms to capitalise on this opportunity. 

\qtop events, especially semileptonic, allow one to trigger on both leptons and jets  independently, 
thereby allowing the possibility to measure trigger efficiencies. 

Moreover, the presence of $W$-bosons,  light jets and $b$-jets 
in \qtop events, allows one to
measure the JES for both light jets and $b$-jets. The 
goal of measuring  
the top-quark mass with an uncertainty of  1 GeV requires a 1\% error on the JES, 
which can be achieved with at least 1/fb of data.

Since almost every decaying top quark produces a $b$ quark, 
\qtop events supply a pure sample of $b$-jets, 
which could be used to calibrate the $b$-taggers. Much work is in progress
to develop these techniques in preparation for the first data-taking.




\begin{table}[hp]
\begin{tabular}{l}
\hline
Isolated $e$ or $\mu$ with $p_{T} > 20$~GeV\\
$\geq$ 4 jets with $p_{T} >40$~GeV\\
2 b-jets\\
\met $>$ 20 GeV\\
\hline
\end{tabular}
\caption[Semileptonic decay offline selection cuts]
{ \small{Offline selection cuts for semileptonic decays.}}
\label{tab:tt_offline_sel}
\end{table}%

\begin{table}[hp]
\begin{tabular}{l}
\hline
Isolated $e$ or $\mu$ with $p_{T} > 20$~GeV\\
$\geq$ 4 jets with $p_{T} > 40$~GeV\\
2 $b$-jets\\
\met $>$~20GeV\\
$M_{jjj} \sim M_{t}$\\
\hline
\end{tabular}
\caption[Offline selection cuts for JES]{ \small{Offline selection cuts for selecting a sample suitable for jet energy scale determination.}}
\label{tab:jes_offline_sel}
\end{table}%

\end{document}